\documentclass[12pt]{article}

\def\greaterthansquiggle{\raise.3ex\hbox{$>$\kern-.75em\lower1ex\hbox{$\sim$}}}
\def\lessthansquiggle{\raise.3ex\hbox{$<$\kern-.75em\lower1ex\hbox{$\sim$}}}
\newcommand{\beq}{\begin{equation}}
\newcommand{\eeq}{\end{equation}}
\newcommand{\beqa}{\begin{eqnarray}}
\newcommand{\eeqa}{\end{eqnarray}}
\newcommand{\beqan}{\begin{eqnarray*}}
\newcommand{\eeqan}{\end{eqnarray*}}
\newcommand{\ba}{\begin{array}}
\newcommand{\ea}{\end{array}}

\newcommand{\A}{{\cal A}}

\def\nz{\ifmmode {I\hskip -3pt N} \else {\hbox {$I\hskip -3pt N$}}\fi}
\def\zz{\ifmmode {Z\hskip -4.8pt Z} \else
       {\hbox {$Z\hskip -4.8pt Z$}}\fi}
\def\qz{\ifmmode {Q\hskip -5.0pt\vrule height6.0pt depth 0pt
       \hskip 6pt} \else {\hbox
       {$Q\hskip -5.0pt\vrule height6.0pt depth 0pt\hskip 6pt$}}\fi}
\def\rz{\ifmmode {I\hskip -3pt R} \else {\hbox {$I\hskip -3pt R$}}\fi}
\def\cz{\ifmmode {C\hskip -4.8pt\vrule height5.8pt\hskip 6.3pt} \else
       {\hbox {$C\hskip -4.8pt\vrule height5.8pt\hskip 6.3pt$}}\fi}

\def\au{{\setbox0=\hbox{\lower1.36775ex%
\hbox{''}\kern-.05em}\dp0=.36775ex\hskip0pt\box0}}
\def\ao{{}\kern-.10em\hbox{``}}

\normalsize
\begin{document}
\bibliographystyle{plain}

\begin{titlepage}
\begin{flushright} UWThPh-2004-9\\

\today
\end{flushright}
\vspace*{2.2cm}
\begin{center}
{\Large \bf  Separability for lattice systems at high temperature}\\[30pt]

Heide Narnhofer  $^\ast $\\ [10pt] {\small\it}
Institute for Theoretical Physics \\ University of  Vienna, Austria\\
\end{center}

\vfill \vspace{0.4cm}

\begin{abstract}Equilibrium states of infinite extended lattice systems at
high temperature are studied with respect to their entanglement.
Two notions of separability are offered. They coincide for finite
systems but differ for infinitely extended ones. It is shown that
for lattice systems with localized interaction for high enough
temperature there exists no local entanglement. Even more
quasifree states at high temperature are also not distillably
entangled for all local regions of arbitrary size. For continuous
systems entanglement survives for all temperatures. In mean field
theories it is possible, that local regions are not entangled but
the entanglement is hidden in the fluctuation algebra.

 \vspace{0.8cm} PACS numbers: 03.65Ud, 03.67Hk, 05.30Fk

\\
\hspace{1.9cm}

\end{abstract}

\vfill {\footnotesize}

$^\ast$ E--mail address: narnh@ap.univie.ac.at
\end{titlepage}

\section{Introduction}
Entanglement of quantum systems means that it is impossible to
describe the state over the combined system only by its local
constituents. This fact has been started to be investigated in
systems built by many constituents, especially for systems over a
lattice $Z^{\nu}$ with $\nu $ the finite dimension of the lattice
in a state invariant under translation \cite{V}, \cite{Vs}. As it
is discussed in \cite{W} translational invariance restricts the
possibility of entanglement between two lattice points. On the
other hand the entanglement of subsystems can increase with the
size of the subsystems essentially to the maximal possible value
\cite{HF}.

In this note we will consider two definitions for separability: on
one hand we can demand that the state is a linear superposition of
product states. On the other hand we can demand that for any two
localized subsystems the entanglement between them vanishes. It
will be shown that for finite systems, i.e. for systems consisting
of a finite number of tensor products, the two definitions
coincide. In the thermodynamic limit however the equilibrium
states of the Ising model that is essentially a classical model
can not be written as linear superposition of tensor product
states if we do not accept measures over the state space that ask
for some care (compare \cite{M}). However the state restricted to
all local subsystems is separable and we will take this fact as
characteristic for the absence of entanglement \cite{N2}.

We will study whether interacting systems at high temperature
become separable. Based on a proof of the uniqueness of
equilibrium states at high temperature by \cite{Gal3}, \cite{BR}
the control, how equilibrium states can be constructed as
perturbation of the tracial state also suffices to prove that the
local entanglement vanishes. However the estimate to find a bound
on the critical temperature above which no entanglement can show
up depends on the size of the subsystems. Therefore non localized
entanglement cannot be excluded.

The above argument does not work for continuous systems. In the
frame work of relativistic quantum field theory the analogue of
the Reeh- Schlieder theorem in temperature states \cite{Jae}
together with the entanglement witness found in \cite{N}
guarantees also for equilibrium states that every two local
regions are entangled. For the nonrelativistic free time evolution
a scaling argument shows that again entanglement survives, but it
becomes more and more localized in space and therefore more and
more energy is needed for its detection. For quasifree lattice
systems however, where the state is determined by the two point
function we have enough control to show, that at least the
entanglement cannot be distillable. It is also hard to imagine how
with exponentially decreasing spacial correlation functions a
delocalized entanglement witness can be constructed.

New features appear in mean field theories as the BCS- model. The
state factorizes over local regions and therefore considered as
state over the quasilocal algebra the state is not entangled in
our sense. If we move however to the algebra of fluctuations
\cite{GVV} correlations between nearest neighbors disappear in
such a weak way that correlations between macroscopic bulks remain
and entanglement can occur for the fluctuation algebra.

\section{Separable states and localized entanglement}

For a finite multipartite system separable states are defined as a
linear combination of tensor product states \beq\omega =\sum
_i\lambda _i \omega ^1_i \otimes ..\otimes \omega ^k_i\quad
\lambda _i\geq 0 \quad \sum _i \lambda _i=1\eeq where $k$ is the
number of constituents. If in the one dimensional situation $\nu
=1$ we order the points $i=1,,k$ on a circle and assume
periodicity, i. e. $\A_{k+l}=\A _l$ and take the shift $\tau
\A_l=\A_{l+1}$ then we can construct translationally invariant
separable states \beq \bar{\omega }=\frac{1}{k}\sum _{l=1}^k\omega
\circ \tau ^l \eeq Generalization to higher dimensions is
obviously possible. However this procedure does not work when
$k\rightarrow \infty$. Separable states , i.e. linear combinations
of product states are only translationally invariant if they are
finite combinations of periodic states so that it is sufficient to
reduce the sum in (2) over the period.

If an equilibrium state is a product state then necessarily
$H=\sum_{i\in Z^{\nu } } ..\otimes 1 \otimes h_i \otimes
1\otimes..$. Therefore there is no entanglement for all
temperatures. Assume we can divide the state into product states
over $\A _{\Lambda }\otimes \tau ^{|\Lambda |}\A_{\Lambda }..$
where $\Lambda =\cup(1,..k), \A _{\Lambda }=\A_1 \otimes \A _2
\otimes..\A_k , |\Lambda |=k<\infty.$ With the same argument
$H=\sum_l \tau ^{kl}h$ for some $h\in \A _{\Lambda }.$ This
forbids interaction between neighbors. Together with the
assumption that the hamiltonian is invariant under space
translation the class of permitted hamiltonians cannot be
generalized. Therefore separable equilibrium states for finite
temperature can only exist if there is no interaction between
lattice points. An exception is the BCS -model, but the effective
hamiltonian in the thermodynamic limit again reduces to a
hamiltonian of the above type \cite{WT}.

The Ising model $H=-\sum _{j,k}J_{j,k}\sigma _z^j \sigma _z^k,
|j-k|=1$ can be considered both as classical and also as quantum
mechanical model, depending on the choice of the quasilocal
algebra, on which $H$ acts. The classical equilibrium state can be
extended to the quantum model. It contains classical spacial
correlations. Therefore for every finite subsystem it is the sum
over product states, but it is not separable in the above sense.
Nevertheless restricted to any local subsystems of finite size the
systems will not be quantum mechanically entangled. This suggests
the definition:

\paragraph{Definition 1:}Let $ \Lambda 1=(i_1,..i_n),|\Lambda1|=n,\Lambda 2=(k_1, ..k_r),|\Lambda 2|=r, i_s\neq k_t.$ A state
is not entangled if restricted to any two local subsystems $\A
_{\Lambda1}\otimes \A _{\Lambda 2}$ it is not entangled.

\paragraph{Definition 2:} A state is not
entangled to order N if for any two local subsystems $\A
_{\Lambda1}\otimes \A _{\Lambda 2}$ with $|\Lambda 1|<N,|\Lambda
2|<N$ the state is not entangled.

\paragraph{Definition 3:}An invariant state is
delocalized entangled of order N if it is entangled in the sense
of Definition 1 but not in the sense of Definition 2, i.e. if  all
algebras $\A _{\Lambda 1}, \A _{\Lambda 2}$ with $|\Lambda 1|<N+1$
or $|\Lambda 2|< N+1$ are not entangled, but there exist algebras
$\A _{\Lambda 1}, \A _{\Lambda 2}$ with $|\Lambda 1|=N+1, |\Lambda
2|= N+1,$ such that $\A_{\Lambda 1} $ and $\A_{\Lambda 2}$ are
entangled.

In this way we classify how entanglement witnesses have to be
localized. Of course it is always possible to take a linear
superposition of entanglement witnesses and in this way to obtain
a new delocalized one. But only if we search for the one that is
localized as much as possible we get information about local
correlations. In fact examples of translationally invariant states
that are delocalized entangled can be constructed, and these
examples are pure states \cite{NH} with weak decay of the spacial
correlations.

In \cite{N2} it was argued that according to the fact, that the
set of entangled states is
 open, a state is either not entangled or delocalized entangled of
 some order $N\geq1.$ This corresponds to the fact, that we
 consider only entanglement witnesses that can be approximated by
 local operators. In  chapter 5 we will extend the definition of
 entanglement to mesoscopic observables and will give an example,
 in which entanglement appears, that is not observable on the
 local level.

It remains to argue that for finite system we have not changed the
definition of separability, that is  that for finite systems every
state that is not entangled to every order is also separable. This
follows from the following observations: the states that
restricted to two subsystems are not entangled and therefore
separable states form a convex set in state space. The
intersection of convex sets is again convex. A convex set is
characterized by its boundary: in every neighborhood of a point of
the boundary there lies a point that does not belong to the set.
Therefore it is not separable with respect to some subalgebras $\A
_{\Lambda }\otimes \A _{\Lambda ^c}$ where we can choose the
second algebra to be the complement, because entanglement is
monotonically increasing with the algebras \cite{N2}. Therefore an
extremal not entangled state is a product state for two
subalgebras. We continue by induction. This state over $\A
_{\Lambda }$ is again not entangled for the subalgebras of
$\A_{\Lambda }$. The boundary of all these states consists of
states that are products of states over some subalgebras of $\A
_{\Lambda}$. We can continue by induction reducing in every step
the size of the algebra. Therefore in a finite number of steps we
reach the algebra over a lattice point. It follows that in fact
the state can be written as linear combination of factorizing
states.

\section{Entanglement at high temperature}
We consider equilibrium states for hamiltonians of the form
$H=\sum _{k\in Z }\tau ^k H_{\Lambda }$ where $\Lambda $ is some
finite region. Since $\Lambda $ is not just a point we permit
interaction between neighboring regions. For simplicity we assume
finite range interaction but the results can be extended for
interactions that decrease sufficiently with the distance (\cite
{BR}). Therefore our considerations are applicable to all popular
examples as the ferromagnetic and antiferromagnetic Heisenberg
model, Hubbard model etc. For all these models the time evolution
$\alpha _t$ is well defined as automorphism on the quasilocal
algebra, i.e. the norm closure of the strictly local operators.
Their equilibrium states satisfy the KMS condition
$$\omega (A B)=\omega (B\alpha _{i\beta}A)$$
with $\beta $ the inverse temperature. For $\beta \rightarrow 0$
we obtain the tracial state, which is separable. That this is not
only formally so but that the equilibrium state can be obtained by
perturbation of the tracial state was shown in \cite{Gal3}. Of
course the perturbation series in $\beta $ has in general finite
convergence range, otherwise phase transitions would be forbidden.
The possibility of perturbation theory guarantees that for high
enough temperature the equilibrium state is close to the tracial
state. We have to specify the way of convergence. We need norm
convergence in an appropriate norm, because the limit must exist
for all operators. But this norm can not be the operator norm
because representations corresponding to states at different
temperature are not equivalent. For our entanglement definition
however we need uniformity for operators that are localized to the
same amount. All these requirements are met by the calculations in
\cite{BR}.

To give a taste of the calculation we repeat the definitions and
estimates following \cite{BR}: With the matrix units
$e(I_X,J_X)=\Pi _{l=i}^n e(i_{x_l},j_{x_l}),$
$X=(x_1,..x_n),I_X=(i_{x_1},..i_{x_n}), J_X=(j_{x_1},..j_{x_n})$
the state is determined by $\bar{\omega }(I_X,J_X)=\omega
(e(I_X,J_X)).$ Let $|\omega |=sup |\omega (e(I_X,J_X))|.$ Let
$$\bar{\delta }(I_X,J_X)=1 \quad for X=\Phi,\quad =\frac{1}{d+1}\delta
_{i_x,j_x} \quad for X=(x),\quad =0 \quad otherwise$$ Here $d$ is
the dimension of the algebra at a lattice point. Then the KMS
condition implies
$$\bar{\omega }=\bar{\delta }+K\bar {\omega } +L_{\beta, H_{\Lambda
}}\bar{\omega }$$ with
$$(Kf)(I_X,J_X)=\frac{1}{d+1}\delta
_{i_{x_1},j_{x_1}}f(I_{X'},J_{X'}), \quad
X=(x_1,...x_n),X'=(x_2,..x_n)$$ and $L_{\beta ,H_{\Lambda }}$ is a
lengthy expression (see \cite{BR}) reflecting the complex time
evolution of the matrix units. We only need the result that
$|K|=\frac{1}{d+1}$ and $|L_{\beta ,H_{\Lambda }}|$ can be bounded
by $(d+1)^2 2\beta |||H_{\Lambda }|||(1-2\beta |||H_{\Lambda
}|||)^{-1}$ where $|||H_{\Lambda }|||$ is some appropriate norm on
the interaction \cite{BR}. With $\bar{\omega }=\sum _n(K+L_{\beta
,H_{\Lambda }})^n\bar{\delta}$ we notice that for sufficiently
small $\beta $ the series converges and  the resulting state will
be close to the tracial state for matrix units, if only $\beta $
is sufficiently small.

Further we know  that in a neighborhood of the tracial state
(whose size $\epsilon _N$ again depends on the dimension of the
algebras ) every state is separable. According to the estimates in
\cite{BR} for every N there exists a critical temperature $\beta
(N,\epsilon ),$ such that $|\omega _{\beta }(A)-tr(A)|<\epsilon
_N||A|| \quad \forall \beta <\beta (N,\epsilon _N), A \in \A
_{\Lambda }, |\Lambda |<N.$  This guarantees the absence of
entanglement for sufficiently localized regions and sufficiently
high temperature, where the critical value for the temperature
depends on the size of the localization and of course on the
details of the interaction.

\section {Entanglement in quasifree states}
The estimates in \cite{Gal3}, \cite{BR} are very general and take
into account that the non commutativity effects increase with
increasing size of the algebra. However we know that in every
extremal equilibrium state the correlations tend to infinity with
increasing distance. Therefore it is plausible to expect that in
general entanglement occurs already for reasonably small
subalgebras or not at all. We will examine this behavior in the
example of Fermi systems on a lattice with a quasifree time
evolution. Though here only the even part of the localized
algebras commute and the odd part has to be handled with care
\cite{Mo} entanglement of even states is not effected. Equilibrium
states are even, therefore we can rely on the characterization of
entanglement offered in \cite{Ho}. Since the time evolution by
assumption is quasifree also the equilibrium states are quasifree.
A generale state can be calculated if we know all expectation
values of $e^{iA}$, where $A$ runs over all selfadjoint operators
quadratic in creation and annihilation operators. Applying the
transposed these operators remain quadratic. For a quasifree state
the expectation value of $e^{iA}$ is in one to one correspondence
to the two point function. Therefore it is sufficient to examine
whether under the application of a partial transposed the two
point function determines a state.

More precisely we consider a hamiltonian $H=\sum _{x,y}a^+_x
V(x-y)a_y$. We can regard $V$ to be an operator in $l^2$. Then the
equilibrium state is given by \beq \omega
(a^*(f)a_(g))=<g|\frac{1}{1+e^{\beta V}}|f>\eeq where $f,g\in
l^2.$ All other expectation values can be written as polynomial
over these two point functions. Especially (we assume $<f|g>=0$)
$$\omega (a^*(f)a(g)+a(\bar{g})a^*(\bar{f})+
a^*(g)a(f)+a^*(f)a(f)
+a(\bar{f})a^*(\bar{f})+a^*(g)a(g)+a(\bar{g})a^*(\bar{g})=$$
\beq\left \langle \ba{c|cccc|c} f  & a_{11}  & a_{12} & 0 &  0  &
f \\ g &  a^*_{12} & a_{22} & 0 & 0 & g
\\ \bar{f}  & 0 & 0 & 1-a_{11} & a^*_{12} &  \bar{f} \\ \bar{g} &
0 & 0 & a^*_{12} & 1-a_{22} & \bar{g}
 \ea \right \rangle \eeq where $a_{11},a_{12},a_{22}$ are
 determined by (3). More precisely
 the matrix in (4) consists of the parts \beq\left( \ba{cc} A &  B
\\ B^*& 1-A
 \ea \right )\eeq where $\omega (a^+(f)a(g))=<g|A|f>,\omega
 (a(g)a^*(f))=<g|1-A|f>, \omega (a(f)a(g))=<\bar{g}|B|f>$
 In a gauge invariant equilibrium state, as it corresponds to our hamiltonian $H$, $B=0$.
 In general $B\neq 0$, but in order that (4) defines a state the above
 expression has to be positive definite which corresponds to
 $A(1-A)\geq B^*B$. If we now apply the transposition only on one
 part, i.e. $ a(f)\rightarrow a(f), a(g)\rightarrow a^*(\bar{g}), a^*(g)a(g) \rightarrow a^*(g)a(g)$
 then (4) becomes \beq\left \langle \ba{c|cccc|c} f  & a_{11}  & 0 & 0 &  a_{12}  &
f \\ g &  0 &  a_{22} & a_{21} & 0 & g
\\ \bar{f}  & 0 & a^*_{21} & 1-a_{11} & 0 &  \bar{f} \\ \bar{g} &
a^*_{12} & 0 & a & 1- a_{22} & \bar{g}
 \ea \right \rangle \eeq
 The positivity requirement becomes \beq a_{11}(1-a_{22})\geq
 a_{12}a^*_{12},\eeq  and the inequality has to hold with respect to all pairs $f,g$. When we vary $f\in H _{\Lambda 1}
 \subset l^2$ and $g\in H _{\Lambda 2} \subset l^2$ where
 $H _{\Lambda 1} $ and $H _{\Lambda 2} $ are orthogonal subspaces
 corresponding to two localized regions then we have to read the equation
  as equation between operators. If the positivity condition
 is violated, then according to the results of \cite{Ho} the state
 between the algebras $\A _{\Lambda 1}$ and $\A _{\Lambda 2}$ is
 entangled, if not then it is at least not distillably
 entangled.
 However $A=\frac{1}{1+e^{\beta V}}$ converges in norm to
 $\frac{1}{2}$ for $\beta \rightarrow 0.$ Therefore (7) is satisfied for sufficiently high
 temperatures. Especially also the [CHSH] inequality for any
 choice of operators cannot serve as entanglement witness.

A similar analysis holds, if we consider bosons in a quasifree
state where  \beq \omega (a^*(f)a_(g))=<g|\frac{1}{e^{\beta V+\mu
}  -1}|f>\eeq Then $$\omega (a^*(f)a(g)+a(\bar{g})a^*(\bar{f})+
a^*(g)a(f)+a^*(f)a(f)
+a(\bar{f})a^*(\bar{f})+a^*(g)a(g)+a(\bar{g})a^*(\bar{g})=$$
\beq\left \langle \ba{c|cccc|c} f  & a_{11}  & a_{12} & 0 &  0  &
f \\ g &  a^*_{12} & a_{22} & 0 & 0 & g
\\ \bar{f}  & 0 & 0 & 1+a_{11} & a^*_{12} &  \bar{f} \\ \bar{g} &
0 & 0 & a^*_{12} & 1+a_{22} & \bar{g}
 \ea \right \rangle \eeq If we take partitions of the algebra
  into account that correspond to (6) then the condition on
  positivity is always satisfied. Entanglement between subalgebras only occurs if
  the partition into the subalgebras includes a non trivial Bogoliubov transformation,
   including a mixture of creation and annihilation operators. But even then in the limit $\beta \rightarrow 0$
 $a_{11},a_{22} \rightarrow \frac{1}{e^{\mu }-1}, a_{12}
 \rightarrow 0$ which again guarantees that for high enough
 temperature right and left region are not entangled.

 The situation is different for continuous systems and vanishing chemical potential. Consider first
 free fermions. Then \beq \omega (a^*(f)a(\bar{g}))=\int dp
 \tilde{f}(p) \tilde{g}(p)\frac {1}{1+e^{\beta p^2}}\eeq
 Since we can observe entanglement between two appropriate modes
 $f,g$
 for $\beta =1$ we can observe entanglement for $\tilde{f}_{\beta }(p)
=\sqrt {\beta }\tilde{f}(\beta p),\tilde{g}_{\beta }(p) =\sqrt
{\beta }\tilde{g}(\beta p).$ If $f$ was localized in the right
$[0,\infty ) $ and $g$ in the left $(-\infty,0]$ also the scaled
functions are localized and entanglement cannot disappear in the
high temperature limit. The same effect occurs in temperature
states in relativistic quantum field theories. Here \cite{Jae} has
shown that the Reeh- Schlieder theorem is still valid, the GNS
vector is cyclic and separating for every local subalgebra. But
this implies entanglement \cite{N}. It means, that the effect of
any local manipulation of the state can also be achieved by
manipulations done in another local region, which is exactly what
entanglement enables.

\section{Delocalized entanglement}
As stated in \cite{N2} on the basis of the quasilocal algebra
entanglement can either be observed already on the local level, if
we only choose the local region sufficiently large, or it cannot
be observed at all. This is in contradiction to the result in
\cite{JKP}, where mesoscopic entanglement was measured. However in
this context the entanglement witnesses have been no quasilocal
operators and also not the mean of quasilocal operators as in
\cite{V}. They belong to the algebra of fluctuations. The precise
definition of this algebra is given in \cite {GVV}. If one defines
an operator $\vec S $ by \beq \hat{\omega }(e^{i{\vec a}{\vec S}
})=\lim _{N\rightarrow \infty}\omega _{\beta}(e^{i{\vec a}\sum
\frac {{\vec \sigma ^k}-\omega ({\vec \sigma ^k}) }{\sqrt N
}})\eeq then the operators satisfy commutation relations similar
to the Weyl algebra \beq\hat{\omega
}(e^{iaS_x}e^{ibS_y})=e^{-iabs_z}\hat{\omega
}(e^{ibS_y}e^{iaS_x}).\eeq where $s_z=\omega (\sigma _z^k)$ in a
state invariant under translations. Therefore we can interpret
$e^{iaS_x},e^{iaS_y}$ as operators in a Weyl algebra, that we call
fluctuation algebra, and $\hat{\omega }$ as state over it. We can
divide the fluctuation algebra into a right and a left part\beq
(e^{i{\vec a}{\vec S}_l })=\lim _{N\rightarrow \infty}(e^{i{\vec
a}\sum_{k=-N \alpha }^0 \frac {{\vec \sigma ^k}-\omega ({\vec
\sigma ^k}) }{\sqrt N }})\eeq \beq (e^{i{\vec a}{\vec S}_r })=\lim
_{N\rightarrow \infty} (e^{i{\vec a}\sum _{k=1}^{(1-\alpha
)N}\frac {{\vec \sigma ^k}-\omega ({\vec \sigma ^k}) }{\sqrt N
}})\eeq and get in this way two commuting Weyl algebras. As in
\cite{N3} we consider the equilibrium states corresponding to the
two hamiltonians\beq H_N^{(1)} = \sum _k ^N a \sigma _z^k\eeq \beq
H_N^{(2)} =\sum _k ^N c \sigma _z^k +\frac{1}{N} \sum _{k,l}^N (
\sigma _x ^k \sigma _x ^l + \sigma _y ^k \sigma _y ^l+\sigma _z ^k
\sigma _z ^l ).\eeq Both hamiltonians lead to the same type of
equilibrium state on the quasilocal algebra, namely \beq \omega
_{\beta } ({\vec \sigma ^k}{\vec \sigma ^l})=\omega _{\beta}
({\vec \sigma ^k})\omega _{\beta} ({\vec \sigma ^l})\eeq with
$\omega _{\beta }(\sigma _z^k)=s_z(\beta ),\omega _{\beta }(\sigma
_x^k)=\omega _{\beta }(\sigma _y^k)=0$, where $a=a(c, \beta )=
c+\omega _{\beta }(\sigma _z^k)$ in (15) is an effective field
strength. If we evaluate the right and the left fluctuation
algebra in this limit state, then the state factorizes and we have
no entanglement. If however we couple the limit in the state with
the limit in the operator \beq \hat{\omega }(e^{i{\vec a}{\vec S}
})=\lim _{N\rightarrow \infty}\frac{1}{Tr e^{- H_N^{(2)}}}Tr e^{-
H_N^{(2)}}(e^{i{\vec a}\sum \frac {{\vec \sigma ^k}-\omega ({\vec
\sigma ^k}) }{\sqrt N }})\eeq then we are considering the limit of
an equilibrium state with respect to a time evolution that
converges in the limit to the one determined by $H_N^{(2)}$ and
differs from the one given by $H_N^{(1)}$: \beq
\frac{d}{dt}S_{rx}=-(c+\frac{2}{N}S_{lz})S_{ry}-\frac{2}{N}S_{rz}S_{ly},
\frac{d}{dt}S_{ry}=(c+\frac{2}{N}S_{lz})S_{rx}-\frac{2}{N}S_{rz}S_{ly}\eeq
\beq\frac{d}{dt}S_{lx}=-(c+\frac{2}{N}S_{rz})S_{ly}-\frac{2}{N}S_{lz}S_{ry},
\frac{d}{dt}S_{ly}=(c+\frac{2}{N}S_{rz})S_{lx}-\frac{2}{N}S_{lz}S_{ry}.\eeq
$\frac{2}{N}S_{rz},\frac{2}{N}S_{lz}$ can be replaced by their
expectation value. Then the evolution becomes linear and is given
by a Hamiltonian of the form\beq H=\nu_1(x^2+p^2)+\nu _2(y^2+q^2),
[x,p]=i,[y,q]=i\eeq with \beq x=a_1S_{rx}+b_1S_{lx},
y=a_2S_{rx}+b_2S_{lx}\eeq where the constants can be calculated
from (19),(20) and are determined by the effective interaction
between right and left. Therefore they depend on $\omega _{\beta
}(\sigma _z^k)$ and thus on the temperature. Analyticity
properties guarantee that also the limit state over the
fluctuation algebra is an equilibrium state with respect to the
limit-time evolution of the fluctuation algebra. More precisely we
can apply Theorem 5.3.12, 6.3.27 and 6.3.28 of \cite {BR}. Since
the time derivative is uniformly bounded we have convergence on
the boundary of the analyticity strip. On this boundary the
hamiltonian corresponds to the quadratic hamiltonian (21) with
expectation value taken over all coherent states. This domain is
sufficient to define the hamiltonian uniquely. Therefore all
conditions of \cite{BR} are met. On the fluctuation algebra the
equilibrium states respectively the groundstate are those of two
harmonic oscillators.

 Since we have two
different rotation velocities the state, which is a Gaussian state
over the two Weyl algebras, does in general not factorize.
Depending on the difference between the two velocities, i. e. on
the temperature, we can examine  entanglement of the right and
left fluctuation algebra on the basis of the results in \cite{N2},
\cite{S}. More explicitly we know that as a consequence of the
Weyl relations \cite{NT} and if $a_1b_1+a_2b_2=0$ (which happens
if $\alpha =\frac{1}{2}$, i.e. if the subalgebras have the same
size) that
$$<S_{rx}^2>+<S_{ry}^2>\quad \geq |\omega(\sigma _z^k)|,<S_{lx}^2>+<S_{ly}^2>\quad \geq |\omega(\sigma
_z^k)|$$ For a separable state this implies
\beq<(S_{lx}+S_{rx})^2>+<(S_{ly}-S_{lx})^2>\geq |\omega (\sigma
_z^k)|\eeq  Inserting the expectation value corresponding to the
temperature this inequality will always hold in the example with
$\alpha =\frac{1}{2}$. Nevertheless also in this situation we have
non trivial classical correlations. This is also in agreement with
the calculation (9). If however we vary the size of the left and
right fluctuation algebra we can violate (23) for small
temperatures. Especially in the groundstate  we have entanglement
on the basis of the fluctuation algebra.\bibliographystyle{plain}

\end{document}